\documentclass[aps,twocolumn,amsmath,amssymb,showpacs,prl,floatfix]{revtex4}
\usepackage{amsmath}
\usepackage{graphicx}
\usepackage{epsfig}
\newlength{\upit}\upit=0.1truein

\newcommand{\ltappr}{{{\lower4pt\hbox{$<$} } \atop \widetilde{ \ \ \ }}}
\newlength{\bxwidth}\bxwidth=1.5 truein

\newcommand{\cg}{{\cal G}}

\newcommand{\str}{\hbox{Str}}

\newcommand{\tr}{{\hbox{Tr}}}

\begin{document}
\newcommand{\dg}{^{\dagger }}
\newcommand{\vk}{\vec k}
\newcommand{\vq}{{\vec{q}}}
\newcommand{\vp}{\bf{p}}
\newcommand{\al}{\alpha}
\newcommand{\be}{\beta}
\newcommand{\si}{\sigma}
\newcommand{\rarrow}{\rightarrow}
\def\fig#1#2{\includegraphics[height=#1]{#2}}
\def\figx#1#2{\includegraphics[width=#1]{#2}}
\newlength{\figwidth}
\figwidth=10cm
\newlength{\shift}
\shift=-0.2cm
\newcommand{\fg}[3]
{
\begin{figure}[ht]

\vspace*{-0cm}
\[
\includegraphics[width=\figwidth]{#1}
\]
\vskip -0.2cm
\caption{\label{#2}
\small#3
}
\end{figure}}
\newcommand{\fgb}[3]
{
\begin{figure}[b]
\vskip 0.0cm
\begin{equation}\label{}
\includegraphics[width=\figwidth]{#1}
\end{equation}
\vskip -0.2cm
\caption{\label{#2}
\small#3
}
\end{figure}}

\newcommand \bea {\begin{eqnarray} }
\newcommand \eea {\end{eqnarray}}
\newcommand{\bk}{{\bf{k}}}
\newcommand{\bx}{{\bf{x}}}

\title{Schwinger Boson approach to the
fully screened Kondo model.
}

\author{
J. Rech$^{1,2,3}$,P. Coleman$^{1,2}$, G. Zarand$^{1,4}$ and O. Parcollet$^{3}$ }
\affiliation{$^{1}$Kavli Institute for Theoretical Physics, Kohn Hall, UCSB
Santa Barbara, CA 93106, USA} 
\affiliation{
$^2$Center for Materials Theory,
Rutgers University, Piscataway, NJ 08855, U.S.A. } 
\affiliation{$^{3}$ SPhT, L'Orme des Merisiers, CEA-Saclay, 91191
Gif-sur-Yvette France.}
\affiliation{$^{4}$Theoretical Physics Department, Budapest University of Technology and Economics,
Budafoki ut 8. H-1521 Hungary }
\begin{abstract}
We apply the  Schwinger boson scheme to the fully
screened Kondo model and generalize the method to include
antiferromagnetic interactions between ions. 
Our approach captures the Kondo crossover from local moment behavior to a Fermi liquid 
with a non-trivial Wilson ratio. When applied to
the two impurity model, 
the mean-field theory describes the "Varma Jones" quantum
phase transition between a valence bond state and a heavy Fermi liquid.
\end{abstract}

%
\maketitle
%


Recently, there has been a broad growth of interest in the
behavior of heavy electron materials at a magnetic quantum
critical point\cite{review}. Several observations 
can not be understood in terms of the established Moriya
Hertz theory of quantum phase transitions\cite{moriya,hertz,millis}, including the divergence of
the heavy electron masses\cite{custers}, 
the near linearity of the
resistivity\cite[]{hilbert1,mathur,malte,steglich} and $E/T$ scaling
in inelastic neutron spectra\cite{schroeder}.
The origins of
this failure are linked to the 
competition between the Kondo effect and antiferromagnetism and
may indicate the emergence of new kinds of excitation\cite{review,si,senthil,pepin05}

In this letter, we show how to 
unify the Arovas Auerbach\cite{arovas}
description of quantum spin systems 
with the physics of the Kondo model by using 
Luttinger Ward
techniques and a Schwinger boson spin representation.
Our goal is to 
develop a large-$N$ expansion that contains the physics of
antiferromagnetism and the Kondo effect in the leading 
approximation.   
Traditional 
pseudo-fermion \cite{abrikosov,read,auerbach} representations of the spin
are ill-suited to a description of 
local moment magnetism. By contrast, a Schwinger boson
scheme works well for magnetism, but to date, has not been 
successfully applied to the 
fully screened Kondo model. We use 
a method co-invented by one of us
(OP)\cite{parcollet97a} in which the 
Kondo effect is captured in a  large $N$ 
Schwinger boson scheme using 
a multichannel Kondo model where the number $K=k N$ 
of screening channels scales
extensively with $N$. 
By tuning the number $n_b$ of Schwinger bosons from $n_{b}<K$ to
$n_{b}>K$, 
one is able to describe both the overscreened and underscreened
Kondo models, however, difficulties were encountered in the past work
that appeared to prevent the treatment of the
perfectly screened case, $n_b=K$.  
In this letter, we show how these difficulties are overcome. 

Consider the multichannel Kondo impurity model,
\begin{eqnarray}\label{l}
\mathcal{H} &=& \sum_{\vk, \nu, \al} \epsilon_{\vk} 
c^{\dg}_{\vk \nu\al} 
c_{\vk \nu \al} 
+ H_{I }- \lambda (n_{b}-2S),\cr
H_{I}&=&
\frac{J_{K}}{N}
\sum_{\nu \al \beta} \psi ^{\dg}_{\nu\al} 
\psi _{\nu \be}  b\dg_{\beta}b_{\alpha }.
\end{eqnarray}
Here $c^{\dg}_{\vk \nu\al} $ creates a conduction electron 
of  momentum $\vk$, channel index $\nu\in[1,K]$, spin index $\alpha \in
[-j,j]$, where $N=2j+1$ is even. $\psi\dg _{\nu \alpha } = \frac{1}{\sqrt{{\cal  N}_{s}}}\sum_{\bk} c\dg
_{\bk \nu\alpha} 
$ creates an
electron in the Wannier state at the origin, where ${\cal N}_{s}$ is the number of sites in
the lattice.  
The operator 
$b\dg _{\alpha}$ creates a Schwinger boson with spin index
$\alpha\in [-j,j ]$. The  local spin operator is represented by 
${S} _{\al \be} = b^{\dg}_{ \al} b_{ \be} - \delta_{\al \be}/N$ and
the system is
restricted to the physical Hilbert space by requiring that 
$n_b = \sum_{\al} b^{\dg}_{\al} b_{ \al} = 2 S$.
The final term in $H$ contains
a temperature-dependent chemical potential $\lambda (T)$
that implements the constraint $\langle n_{b}\rangle =2S$. 
We will examine the fully screened case $2S=K$, taking 
the large $N$ limit where $N\rightarrow \infty $ keeping $k=K/N$ fixed.

We begin by factorizing the interaction in
terms of auxilliary Grassman field $\chi_{\nu} $, 
\begin{equation}\label{}
H_{I}
\rightarrow 
\sum_{\nu\al}
\frac{1}{\sqrt{N}}\left[( \psi \dg_{\nu\alpha }b_{\alpha} )
\chi\dg _{\nu}
+{\rm  H.c.} 
\right] + \sum_{\nu}\frac{\chi\dg _{\nu}\chi_{\nu}}{J_{K}}.
\end{equation}
Following the steps outlined by us in
earlier work\cite{indranil2}, we now write 
the Free energy as a Luttinger Ward\cite{luttinger2} functional  of the one particle Green's functions, 
\begin{equation}\label{notsobigdeal}
F[\cg ] = T\ {\str}\left[\ln  \left(- {\cal G}^{-1} \right)+
(\cg_{0}^{-1}-\cg^{-1}){\cal
G} \right]+ Y[{\cal G}]
\end{equation}
where 
 $\str[A] = \tr[A_{B}]-\tr[A_{F}]$ is  the graded (super) trace 
over the Matsubara 
frequencies, internal quantum numbers of the 
bosonic (B) and fermionic (F) components of $A$. 
$\cg_{0} $ is the bare propagator and $\cg={\rm  Diag}[ 
\underline{G}
_{b}
, \ 
\underline{G}
_{\chi }
, \ 
\underline{G}
_{c}
]
$,
the fully dressed propagator, where 
\begin{eqnarray}\label{a}
\underline{G}
_{b} (i\nu_n)
&=& [{i\nu_n+\lambda -\Sigma_{b} (i\nu_n)}]^{-1}\delta_{\sigma \sigma '},\cr
\underline{G}
_{\chi } (i\omega_n)&=& [{- J_{K}^{-1}-\Sigma_{\chi } (i\omega_n)}]^{-1}\delta_{\nu \nu'},\cr
\underline{G}
_{c}(i\omega_n)&=& [{G_{c0}^{-1} (i\omega_n) -\Sigma_{c} (i\omega_n)}]^{-1}\delta_{\sigma\sigma '}\delta_{\nu \nu'}.
\end{eqnarray}
$\Sigma
(i\omega_n)= \cg_{0}^{-1}-\cg^{-1}
$ denotes the corresponding self-energies.
$G_{c0} (i\omega_n)= \sum_{k}\frac{1}{i\omega_n-\epsilon_{k}}$ is the
bare conduction electron Green's function. 
The quantity
$Y[\cg]$ is the sum of all closed-loop two-particle
irreducible skeleton Feynman diagrams. In the large $N$ limit, we 
take the leading $O (N)$ contribution to $Y$ (Fig. \ref{fig1}).

\figwidth 0.9\columnwidth
\fg{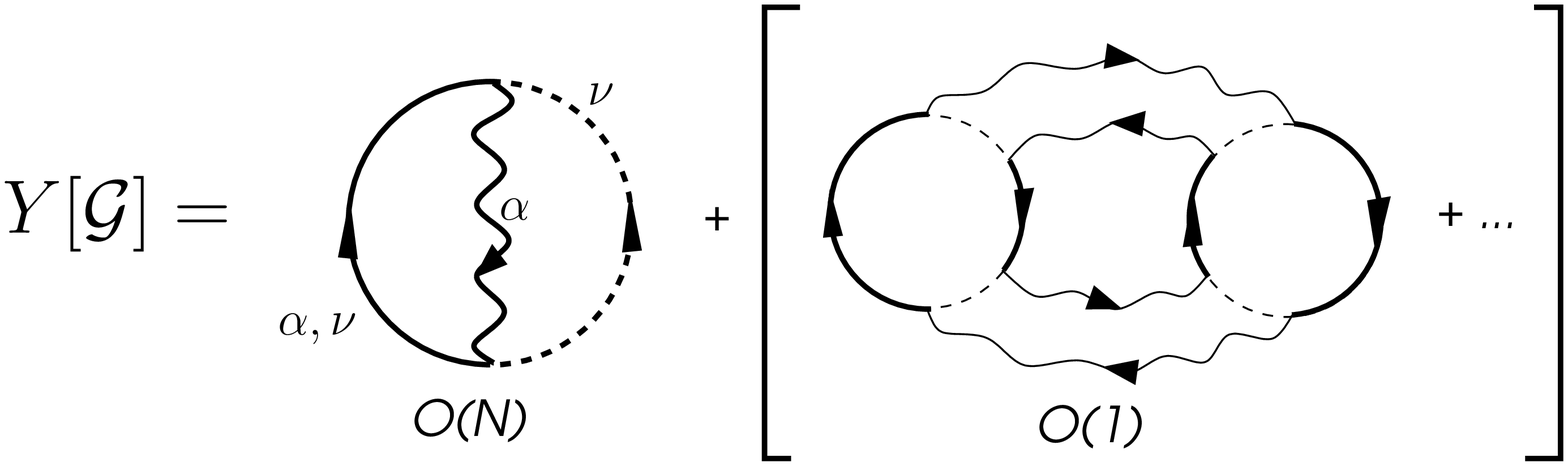}{fig1}{Leading contributions to $Y[\cg ]$ in $1/N$
expansion. Solid, dashed and wavy lines respectively 
represent $G_{c}$, $G_{\chi }$ and $G_{b}$. Each vertex is associated with a factor
$i/\sqrt{N}$. Bracketed terms
are dropped in the large $N$ limit.}

The variation of $Y$ with respect to $\cg  $ generates the self-energy 
$\delta Y/\delta \cg = \Sigma $, which yields 
$
\Sigma_{c} (\tau)=  \frac{1}{N}G_{\chi} (-\tau)G_{b} (\tau )
$.
Since $\Sigma_{c} $ is of order $O (1/N)$ we can use the 
bare conduction propagator 
$G_{c0}$ inside the self-consistent equations for
$\Sigma_{\chi} (\tau)=   G_{b} (\tau)G_{c0} (-\tau)$ and 
$\Sigma_{b} (\tau)= -k G_{\chi} (\tau)G_{c0} (\tau)$.
In terms of real frequencies, these expressions become
\begin{eqnarray}\label{selfc}
\Sigma_{\chi} (\omega-i\delta)&=& 
-\int \frac{ d\nu}{\pi}\left[h_{F}
(\nu)G_{c0}'' (\nu)G_{b} (\omega+ \nu) \right.\cr
&+& \left. h_{B} (\nu)G_{b}''(\nu)G_{c0}^{*} (\nu- \omega) \right]\cr
\Sigma_{b} (\nu-i\delta )&=&k\int \frac{ d\omega}{\pi}
h_{F}(\omega)\left[G_{c0}'' (\omega)G_{\chi } (\nu- \omega) \right. \cr
&+& \left. G_{\chi }''(\omega)G_{c0} (\nu-\omega ) \right]
\end{eqnarray}
where 
$G (\omega)\equiv  G (\omega-i\delta )$ and the primed variables
denote the real and imaginary parts of
$G$. We use the notation $h_{B,F} (\omega)=  \frac{1}{2}\pm n_{B,F} (\omega)
$ where $n_{B,F} (\omega) =
(e^{\be\omega}\mp 1)^{-1}$  are the Bose and Fermi occupation numbers.
The 
constraint $n_{b}=2S$ now becomes
$\int \frac{d\nu}{\pi}n_{B} (\nu) G''_{b}(\nu)  = \frac{2S}{N}$.

The original work in \cite[]{parcollet97a,parcollet2} focussed primarily on the case of the overscreened Kondo
model, where $K>2S$.  The perfectly screened case where $K=2S$
presented two difficulties.
First, the phase shift associated with the
Kondo model is $\pi/N$, which 
vanishes  in the large $N$ limit. Second, 
the requirement that $K=2S$ appeared  extremely stringent, 
the slightest deviation from this condition apparantly leading
to  underscreened or overscreened behavior at low temperatures.

There are two new observations that enable
us to now avoid these difficulties. 
First, the perfectly screened  case where $2S=K$ is 
a stable ``filled shell'' singlet configuration of the
spins. In strong-coupling, this stable ground-state
corresponds to a  singlet ``rectangular'' Young tableau representations of $SU (N)$.
\cite[]{andrei1,andrei2}. 
In the gauge theory description of the Kondo model,  
this stability manifests itself as a gap in the Schwinger boson
and $\chi $ fermion spectrum. 
When the chemical potential, 
$\lambda$ lies within this gap, the ground-state Schwinger boson occupancy
locks into the value $n_{b}=K=2S$ (Fig. 2a). 
\figwidth 1.05 \columnwidth
\fg{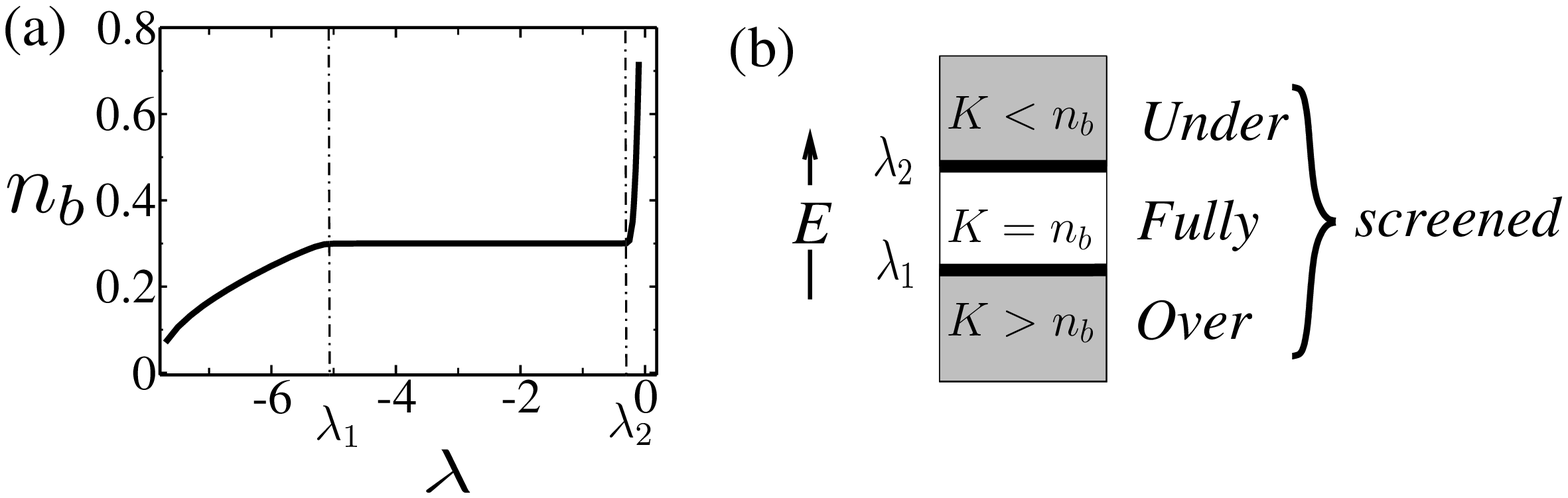}{fig2}{(a) Schematic variation of occupancy as a
functional of boson chemical potential $\lambda$ in the ground-state.
(b) Gap in spinon spectrum develops in the fully screened Kondo state.
}

The gap in the gauge particle spectrum 
has the effect of ``confining'' these excitations, so that the low
energy physics only involves the elastic scattering of the conduction
electrons off the Kondo singlet.  
From this perspective, the fully screened Kondo model is a kind of
 ``spinon insulator'' in this large-$N$ description, while over and 
underscreening develop 
when the spin-chemical potential is in the ``valence'' and ``conduction''
bands respectively ( Fig. 2 (b)). 
A second aspect to the problem concerns the scattering phase shift. 
Although 
the conduction phase shift is $\pi/N$, its  effect on 
the thermodynamics is enhanced by
the  $N$ spin components and the $K$ scattering channels, producing 
an order $O (N\times K/N)\equiv O (N)$
effect in the mean field theory.  Moreover, 
we can identify exact Ward identities\cite{indranil2} which give
rise to a sum rule
relating the scattering phase shift of the conduction electrons
$\delta_{c}={\rm Im}\ln\bigl [ 1 - {g}_{c0} (0-i\delta ) \Sigma_{c}
(0-i\delta )) \bigr]$ to the phase shift
$\delta_{\chi }= {\rm Im}\ 
\ln(1+J_{K}\Sigma_{\chi } (0-i \delta)) 
$ associated with the $\chi $ fermions,
$\delta_{c}= \frac{1}{N}\delta_{\chi }$.
The confined nature of the $\chi $ fermions means that 
$\delta_{\chi }= \pi$ in the ground-state, which 
guarantees that the Friedel sum rule $\delta_{c}= \pi/N$ is
satisfied in this Schwinger boson scheme. 

In the large $N$ limit the entropy\cite{indranil2} is given by
\begin{widetext}
\begin{equation}
\frac{S (T)}{N} = \int \frac{d\omega}{\pi} \left\{ 
\frac{d n_{B}(\omega)}{dT}
 \left[{\rm Im}\ln (-G_b^{-1} ) + G'_{b}\Sigma''_{b} \right] + k\frac{d n_{F}(\omega)}{dT} 
\left[{\rm Im}\ln (-G_{\chi}^{-1} ) + G'_{\chi }\Sigma''_{\chi } - G''_{c 0}\tilde{\Sigma}'_{c}  \right] \right\}
\end{equation}
(where  the frequency labels $\omega$ in the integrand have been
suppressed), and 
\begin{eqnarray}
\tilde{\Sigma}_{c} (\omega-i \delta) =N{\Sigma}_{c} (\omega-i\delta) = 
-\int \frac{ d\nu}{\pi}\left[h_{F}
(\nu)G_{\chi }'' (\nu)G_{b} (\omega+ \nu) \right. 
 + \left. h_{B} (\nu)G_{b}''
(\nu)G_{\chi }^{*} (\nu- \omega) \right]
\end{eqnarray}
is the rescaled conduction self-energy.
At low temperatures, the gap in 
the boson and $\chi $ fermion spectrum
means that only the conduction electron contribution 
dominates the entropy, and this is the origin of the Fermi
liquid behavior. 
\end{widetext}
We can also calculate the local magnetic
susceptibility \begin{equation}\label{}
\chi_{\rm loc} (T) = -2 N \int\frac{d\omega}{\pi} h_{B} (\omega) G'_{b} (\omega)G''_{b} (\omega)
\end{equation}
where we have taken the magnetic moment of
the local impurity to be $M= \sum_{\sigma} \tilde{\sigma }b\dg
_{\sigma}b_{\sigma}$, where $\tilde{\sigma }= {\rm sign}(\sigma)$. 
Note in passing that the dynamic counterpart $\langle S(t) S(0) \rangle$ vanishes exponentially due to the gap in the bosonic spectral functions, the $1/t^2$ term characteristic of a Fermi liquid only appearing at the next order in $1/N$.

\figwidth 0.7 \columnwidth
\fg{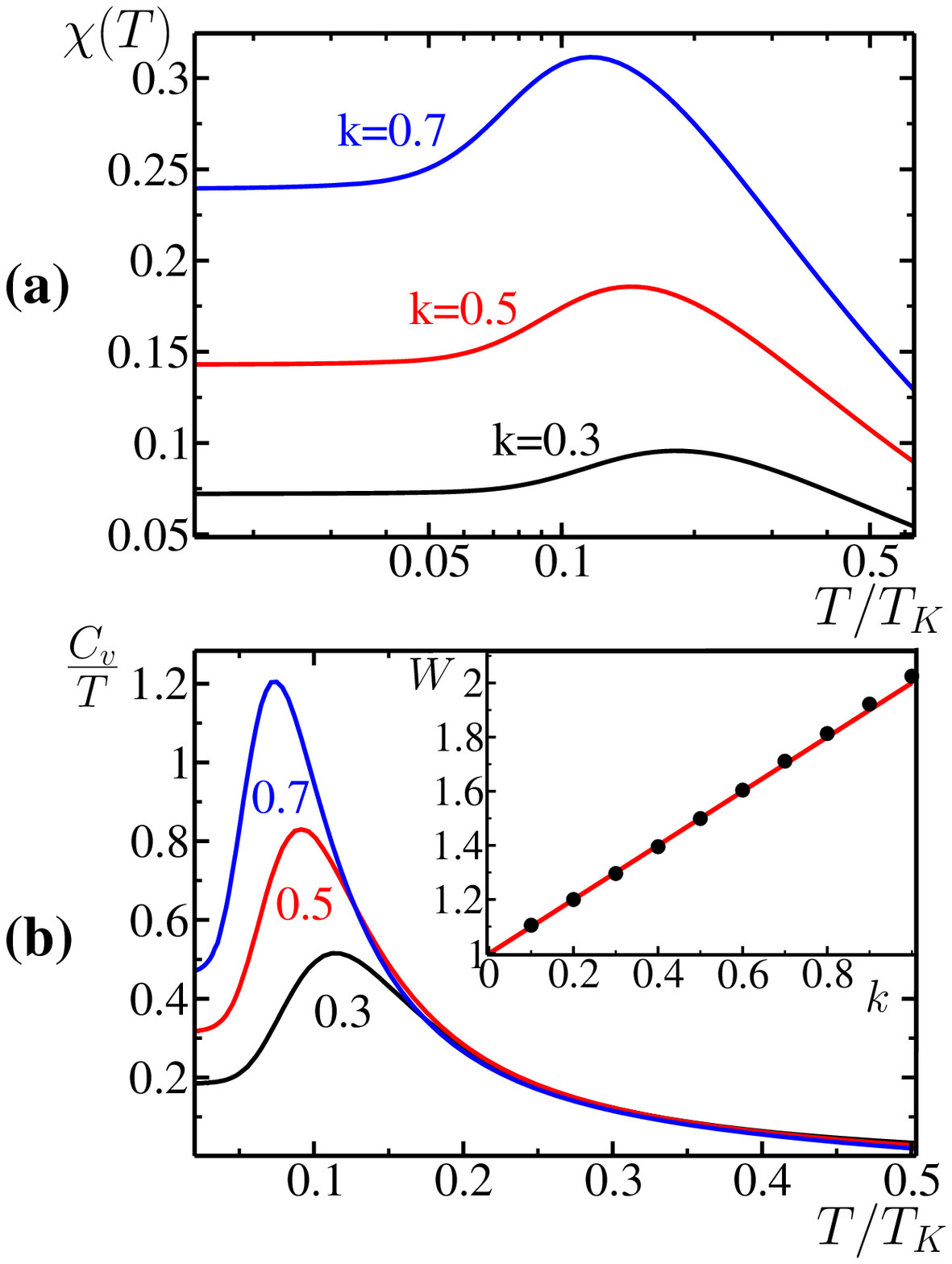}{fig3}{Showing temperature dependence of (a)
impurity magnetic
susceptibility and (b) specific heat capacity in the fully screened
Kondo model for $k=0.3,\ 0.5,\ 0.7$ (with the Kondo temperature given by 
$T_K = D e^{-2D/ J_K}$, $D$ being the electron bandwidth). Inset calculated Wilson ratio. 
}

We have numerically solved the self-consistent equations (\ref{selfc}) for the
self-energy by 
iteration, imposing the constraint at each temperature. 
Fig. 3. shows
the temperature dependent  specific heat coefficient $\frac{C_{V}}{T} =
\frac{dS}{dT}$ and the {\it full} magnetic susceptibility $\chi(T)$.
There is a smooth cross-over from local
moment behavior at high temperatures $\chi \sim n_{b} (1+n_{b})/T$, 
to Fermi liquid behavior at low temperatures.
From a Nozieres-Blandin description of the local Fermi liquid\cite{blandin}
(where channel and charge susceptibility vanish), we deduce the Wilson ratio
\begin{equation}\label{}
W=\frac{\chi / \gamma}{\chi_0 / \gamma_0 }= \frac{ (1+k)}{1 - \frac{1}{N^{2}}}.
\end{equation}
This form is consistent with Bethe Ansatz
results\cite[]{andrei1}. 
We may also derive
this result by applying  Luttinger-Ward techniques\cite{yoshimori}
to our model.  Our large $N$ approximation 
reproduces the limiting large $N$ behavior
of this expression, $W= 1+k$, in other words, the local Fermi
liquid is interacting in this particular large $N$ limit. 


To see how our method handles magnetic correlations, we have
applied it to a two-impurity Kondo model,
\begin{equation}\label{}
H =  \sum_{\vk, \nu, \al} \epsilon_{\vk} 
c^{\dg}_{\vk \nu\al} 
c_{\vk \nu \al} + H_{K } (1) + H_{K } (2) - \frac{J_{H}}{N}B\dg
_{12}B_{12}, 
\end{equation}
where $H_{K }(i)$ is the Kondo hamiltonian for impurity $(i)$
and the antiferromagnetic interaction between the two moments
is expressed in terms of the  boson pair operator
$B_{12}=\sum_{\sigma } \tilde{\sigma }
b_{1\sigma}b_{2 -\sigma }
$.
$H$ is invariant under spin transformations in the  
symmetry group SP (N) (N-even)\cite{readsachdev91}.
We now factorize the antiferromagnetic interaction\cite{arovas},
\begin{equation}\label{}
- \frac{J_{H}}{N}B\dg _{12}B_{12}\rightarrow \bar \Delta  B_{12} +
B\dg _{12}\Delta + \frac{N\bar  \Delta\Delta }{J_H}.
\end{equation}

Boson pairing is associated with the 
establishment of short-range antiferromagnetic correlations. Once
$\Delta $ becomes non-zero, the local gauge symmetry is broken, and the
Schwinger bosons  propagate from site to site. In this ``Higg's
phase'' 
the $\chi $ fermions also delocalize, giving rise
to a mobile, charged yet spinless
excitation that is gapped in the Fermi liquid. Loosely speaking, these
excitations are mobile Kondo singlets or ``holons''.
Since the paired Schwinger bosons interconvert from particle to hole
as they move, they only induce holon motion within the same sublattice.
In the special case of
two-impurity model, so long as the net coupling between the spins
is antiferromagnetic, the holons will remain localized. 
\figwidth 0.98\columnwidth
\fg{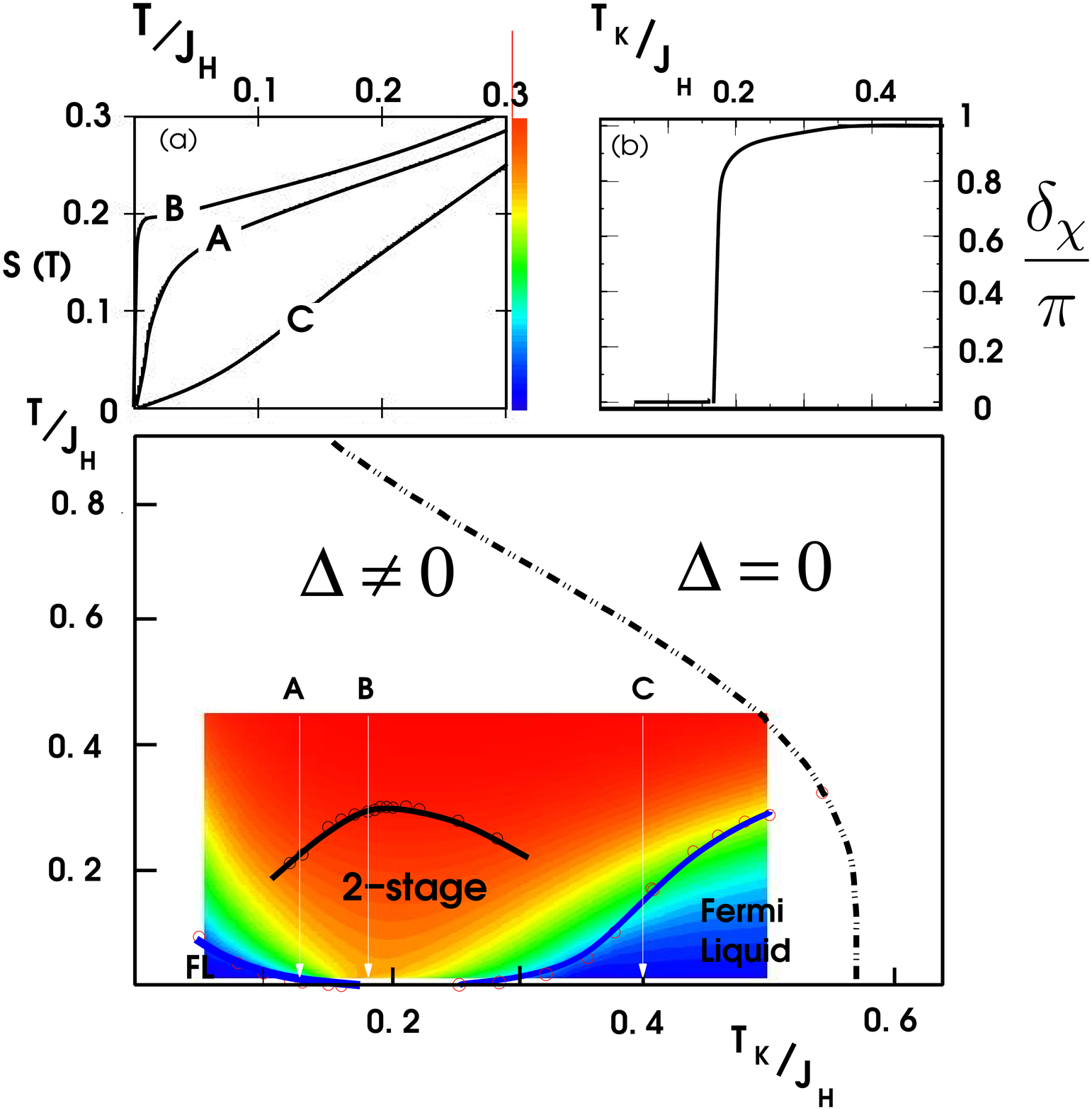}{fig4}{Phase diagram for the two-impurity Kondo model showing
the boundary where boson pairing develops. Color coded contours
delineate the entropy  around the Varma Jones fixed point. Black line
indicates upper maximum in specific heat, blue line, lower maximum in
specific heat where  cross-over into the Fermi liquid takes place. 
Inset - (a)
entropy for various values of $T_{K}/J_{H}$ (b)
showing dependence of $\delta_{\chi }$ on $T_{K}/J_{H}$ at a
temperature $T/T_{K}= 0.02$.
}

Under these assumptions, we can adapt the single impurity 
equations to the two-impurity model by replacing
\begin{equation}\label{}
G_{b} (\omega)\rightarrow \tilde{G}_{b} (\omega)= \left ( G_{b} (\omega)^{-1}
 - \vert \Delta \vert^{2}G_{b} (-\omega)^*
\right )^{-1}
\end{equation}
in the integral equations.
We must also impose 
self-consistency $\Delta = - J_{H}\langle B_{12}\rangle
$, or 
\begin{equation}\label{}
\frac{1}{J_{H}}= - 2 \int \frac{d\omega}{\pi} h_{B} (\omega) {\rm Im} 
\left(
\frac{1}{G_{b}^{-1} (\omega) G_{b}^{-1} (-\omega)^{*} - \vert \Delta 
\vert^{2}} \right)
\end{equation}
We have
self-consistently solved the integral equations 
with the modified boson propagator.
Using the entropy as a
guide, we are able to map out the large $N$ phase diagram for this
model(Fig.\ 4.).

We find that the development of $\Delta\neq  0$
preserves the linear temperature dependence of the
entropy at low temperatures, indicating Fermi liquid behavior.
However, as the $J_{H}$
increases,  the temperature range of Fermi liquid behavior 
collapses towards zero, vanishing
at a quantum critical point where  $J_{H}=J_{c}$. 
For $J_{H}>J_{c}$, 
Fermi liquid behavior re-emerges, but the phase 
shift $\delta_{\chi }$ is found to have jumped from $\pi$ to
zero, indicating a collapse of the Kondo resonance. 
The entropy develops
a finite value at the quantum critical point which is numerically
identical to one half the high temperature entropy of a local moment,
$(N/2)[(1+n_{b})\ln (1+n_{b})- n_{b}\ln n_{b}]$.  
Similar behavior occurs at the ``Varma Jones'' fixed
point\cite{varma,jones,gan} 
in the $N=2$
two-impurity model when the conduction band is particle-hole symmetric.
We can in fact 
identify two maxima in the specific heat, indicating that as in the $N=2$
Varma Jones fixed point, the antiferromagnetic coupling generates a
second set of screening channels, leading to 
a two-stage quenching process. 

The survival of the Varma Jones fixed point at large $N$ in the
absence of particle hole symmetry is a consequence 
of the two impurity Friedel sum rule which tells us that 
\begin{equation}\label{}
(\delta_{+}+ \delta_{-}) = \frac{2\pi}{N}
\end{equation}
where $\delta_{\pm}$ are the even and odd parity scattering phase
shifts. For $N=2$, this condition is  satisfied with $\delta_{+}=
\pi$ and $\delta_{-}= 0$, so it is possible, in the presence of
particle-hole asymmetry,  to cross smoothly 
from unitary scattering off both impurities, to no scattering off
either, while preserving the sum rule. 
However, for $N>2$, where $\delta_{+}+\delta_{-}< \pi$, the sum rule
can not be satisfied in the absence of scattering, 
so the collapse of the Kondo effect must occur via a critical point.

In conclusion, we have shown that a Schwinger boson approach
to the fully screened Kondo model can be naturally 
extended to incorporate magnetic interactions. 
One of the interesting new elements 
is the appearance of mobile, yet gapped ``holon'' excitations in the
antiferromagnetically correlated Fermi liquid.  Future work will
examine whether these excitations can become gapless at a heavy 
electron quantum critical point, leading to quantum
critical matter with spin-charge decoupling\cite{review,pepin05}. 

We are grateful to Matthew Fisher, Kevin Ingersent, Andreas Ludwig and
Catherine P\' epin for discussions related to
this work. 
This research was supported by the
National Science Foundation grant  DMR-0312495 , INT-0130446 and
PHY99-07949. J. R. and O. P. are supported by an ACI grant of the French 
Ministry of Research. 
The authors would like to thank the hospitality of the KITP, where
part of this research was carried out. 



\begin{thebibliography}{99}

\bibitem{review}P. Coleman, C. P\'epin, Q. Si and R.  Ramazashvili,.
{\em {J.\ Phys.: Condens.\ Matter}}{ \bf 13}, R723--R738 (2001).

\bibitem{moriya}T. Moriya and J. Kawabata, J. Phys. Soc. Japan {\bf
34}, 639 (1973); J. Phys. Soc. Japan {\bf 35},669 (1973).

\bibitem{hertz} J. A. Hertz, Phys. Rev. B {\bf 14}, 1165 (1976).

\bibitem{millis} A. J. Millis, Phys. Rev. B {\bf 48}, 7183 (1993).

\bibitem{custers}J. Custers et al, Nature, 424, 524-527 (2003)


\bibitem{hilbert1}H. v. Lohneysen et al. Phys. Rev Lett {\bf 72}, 3262 (1994).

\bibitem{mathur}N. Mathur {\it et al.} Nature {\bf 394},39 (1998).

\bibitem{malte}M. Grosche {\it et al.}, J. Phys. Cond Matt {\bf 12}, 533(2000).

\bibitem{steglich} O. Trovarelli et al, 
Phys. Rev. Lett. {\bf 85 },626 (2000).

\bibitem{schroeder}A. Schroeder {\it et al.}, Nature {\bf 407}, 351(2000).

\bibitem{si} Q. Si, S. Rabello, K. Ingersent and J. L. Smith, Nature 
{\bf 413}, 804 (2001).

\bibitem{senthil} T. Senthil, S. Sachdev and M. Vojta, Physica B,
359-361, 9-16 (2005).

\bibitem{pepin05}C. P\' epin, Phys. Rev. Lett. 94, 066402 (2005)

\bibitem{readsachdev91}N. Read and S. Sachdev, Phys. Rev. Lett,
{\bf  66}, 1773 (1991).








 







\bibitem{arovas}D. P. Arovas,  and A. 
Auerbach,  Phys. Rev {\bf B 38}, 316, (1988).

\bibitem{abrikosov}A. A. Abrikosov,    Physics {\bf 2}, 5 (1965).

\bibitem{read}N. Read, and D. M. Newns, J. Phys. C {\bf 29}, L1055, (1983).

\bibitem{auerbach}A. Auerbach, and K. Levin, Phys. Rev. Lett. {\bf  57}, 877, (1986).

\bibitem{parcollet97a}O. Parcollet and A. Georges, PRL {\bf 79}, 4665-8
(1997); O. Parcollet, A. Georges, G. Kotliar, and A. Sengupta
Phys. Rev. B {\bf 58}, 3794 (1998).

\bibitem{parcollet2}O. Parcollet, PhD. Thesis, unpublished (1998) (http://www-spht.cea.fr/pisp/parcollet/).


\bibitem{blandin}P. Nozi\`eres and A. Blandin, J.  Physique, 41, 193, 1980.

\bibitem{andrei1}A. Jerez, N. Andrei and G. Zarand, Phys. Rev. B58, 3814-41, (1998).

\bibitem{varma}B. A. Jones and C. M. Varma, Phys. Rev. Lett. 58, 843 (1987)

\bibitem{jones}B. A. Jones, B. G. Kotliar, and A. J. Millis, Phys. Rev. B 39, 3415 (1989).

\bibitem{luttinger2}J. M. Luttinger and J. C.  Ward, Phys. Rev. {\bf 118},
1417 (1960).

\bibitem{indranil2}P. Coleman, I. Paul and J. Rech, cond-mat/0503001
to be published (2005).

\bibitem{andrei2}N. Andrei and P. Zinn Justin, Nucl. Phys. B528, 648 (1998).

\bibitem{yoshimori}A. Yoshimori, Prog. Theo. Phys 55, 67-80 (1976).

\bibitem{gan}J. Gan, Phys. Rev. Lett. {\bf 74} 2583-6, (1995).

\end{thebibliography}
\end{document}